\documentclass[useAMS,usenatbib]{mn2e}
\usepackage{latexsym,graphicx,natbib}

\newcommand{\be} {\begin{equation}}

\newcommand{\ee} {\end{equation}}

\newcommand{\RXTE}{{\em R}XTE}
\newcommand{\bc}{\begin{center}}
\newcommand{\ec}{\end{center}}
\def\ltsima{$\; \buildrel < \over \sim \;$}
\def\simless{\lower.5ex\hbox{\ltsima}} 
\def\loe{\lower.5ex\hbox{\ltsima}}
\def\gtsima{$\; \buildrel > \over \sim \;$}
\def\simgreat{\lower.5ex\hbox{\gtsima}}
\def\goe{\lower.5ex\hbox{\gtsima}}

\def\ergscm2 {erg\,s$^{-1}$cm$^{-2}$}
\def\msun {M_{\odot}}

\title[The AMXP Swift J1756.9-2508]{The long-term evolution of 
the accreting millisecond X-ray pulsar Swift J1756.9-2508}

\author[Patruno, Altamirano \& Messenger]
{
Alessandro Patruno$^1$, Diego Altamirano$^1$ \& Chris Messenger$^2$ \\
$^1$Astronomical Institute `A. Pannekoek', Univeristy of Amsterdam,
Science Park 904, Amsterdam, The Netherlands: \tt{a.patruno@uva.nl}\\
$^2$Albert Einstein Institute for Gravitational Research,
  Callinstra{\ss}e 38, 30167 Hannover, Germany }

\begin{document}

\pubyear{2009}

\maketitle

\label{firstpage}

\begin{abstract}
We present a timing analysis of the 2009 outburst of the accreting
millisecond X-ray pulsar Swift J1756.9-2508, and a re-analysis of the
2007 outburst. The source shows a short recurrence time of only $\sim
2$ years between outbursts. Thanks to the approximately 2 year long
baseline of data, we can constrain the magnetic field of the neutron
star to be $0.4\times 10^{8}$G$\simless B\simless 9\times 10^{8}$G,
which is within the range of typical accreting millisecond
pulsars. The 2009 timing analysis allows us to put constraints on the
accretion torque: the spin frequency derivative within the outburst
has an upper limit of $|\dot{\nu}|\simless 3\times
10^{-13}\rm\,Hz\,s^{-1}$ at the 95\% confidence level. A study of
pulse profiles and their evolution during the outburst is analyzed,
suggesting a systematic change of shape that depends on the outburst
phase.

\end{abstract}


\section{Introduction}\label{Intro}

The accreting millisecond X-ray pulsar (AMXP) Swift J1756.9-2508
(henceforth referred to as J1756) was discovered in 2007 by the Burst Alert
Telescope (BAT) aboard the \textit{Swift} satellite \citep{kri07a}.
Follow up observations were performed with the \textit{RXTE}
Proportional Counter Array (PCA) which revealed an X-ray pulsar with a
spin frequency of 182 Hz orbiting in 54.7 min around
a very low mass companion of minimum mass 0.0067$\msun$
\citep{kri07b}.

The X-ray lightcurve of the 2007 outburst showed a very fast 
rise of less than 1 day, followed by an exponential decay 
with an e-folding time of $7.4\pm 0.4$ days.
An inspection of archival \textit{RXTE} All-Sky Monitor (ASM) data 
revealed a lack of detections in the previous 10 years of observations, 
which suggested a long recurrence time for the outbursts. 
However, due to the low flux of the source and the poor sensitivity of the 
ASM, the recurrence time
was only tentative and a shorter duty cycle was not firmly excluded. 

In July 2009 a new outburst was discovered by both \textit{Swift} BAT
and \textit{RXTE} PCA observations \citep{pat09}.
Follow-up observations with the \textit{Swift} X-ray telescope (XRT) 
and  \textit{RXTE} PCA were promptly scheduled and monitored the 
outburst of J1756 until it returned into quiescence. 

In this Letter we report a first analysis of the 2009 outburst
and a re-analysis of the 2007 data to evaluate the long-term 
evolution of J1756 on a baseline of approximately 2 years.

\section{X-ray observations and data reduction}

We used all the \textit{RXTE} PCA data (2007 and 2009) and the 2009
\textit{Swift}-XRT observations for J1756 (Table 1).  We refer to
\citet{jah06} for PCA characteristics and \textit{RXTE}
absolute timing.  We used all available Event and GoodXenon data,
rebinned to 1/8192 s. For the timing analysis, the photons were
extracted from the 2.5-16 keV band to maximize the signal-to-noise
ratio (S/N). We barycentered the photons time of arrivals with the
FTOOL faxbary by using the source positions as determined from the
\textit{Swift}-XRT 2007 observations \citep{kri07b}.  We folded
approximately 500 s long data chunks, keeping only those with S/N$>3$.

The fractional sinusoidal amplitude of the i-th pulse profile
is calculated as:
\begin{equation}\label{fa}
R_{i}= \frac{A_{i}}{N_{ph,i}-B_{i}}
\end{equation}
where $\rm N_{\it ph,i}$ and $\rm B_{\it i}$ are the total number of
photons and the background counts (calculated with the FTOOL
\emph{pcabackest}) in the i-th pulse profile, and $A_{i}$ is the
amplitude (in photon counts) of the observed pulsations.  The error on the
fractional amplitude $R_{i}$ is calculated propagating the errors on
$A_{i}$ and $N_{\it ph,i}$.  The error on $B_{i}$ is negligible with
respect to the other errors and will not be considered further. The
X-ray lightcurve is constructed by using the counts in PCA Absolute
channels 5-37 ($\approx 2.5-16$ keV).

During the 2009 outbursts, the {{\it Swift}-XRT} observed the faint
tail of the outburst with 6 pointed observations (see Table 1)
covering a total of $\sim 11$ ks. All the data were reduced by using
the XRT pipeline (v. 0.12.0). Each observation lasted between 1 and
2.2 ks.  The data were collected in photon counting (PC) mode and the
source and background events were extracted in the energy range
0.5--10 keV, by using circular regions with radii of 20 arcseconds.
Given the presence of a very bright source in the proximity of J1756
(GX 5-1) the first four {{\it Swift}-XRT} pointings were severely
contaminated by the "stray-light" effect \citep{mor09} which limited
the sensitivity of the observations by increasing the background
contamination.

For the spectral analysis we extracted PCA and HEXTE spectra from the
standard modes following the procedures suggested by the RXTE
team\footnote{http://heasarc.gsfc.nasa.gov/docs/xte/xtegof.html},
using the latest available version of the analysis tools by September
1st, 2009.
For the PCA we only used data of PCU 2, which was active in all
observations. For HEXTE we only used data from Cluster B.
No energy channels were binned  in any case. No systematic errors were
added to  HEXTE data,  while we  used the standard  0.5\% for  the PCA
data.  We  fitted simultaneously the PCA  and HEXTE data  in the 3--25
and  25--150 keV  range, respectively.  
A single absorbed blackbody or a single power law component did not fit well the 
data,  so we used  a  standard absorbed
\textit{wabs} \citep{Morrison83}  three-components         model
(disc-blackbody, a  power law and a  Gauss line if  necessary). To fit
the data we used Xspec v12.5.0 \citep{Arnaud96}. 
We tried also to fit
\textit{compTT}   \citep{tit94}   and  \textit{compPS}   \citep{pou96}
spectral models. In all cases we obtained a satisfactory 
fit with reduced $\chi^{2}\sim 0.8-1.01$ for 82 or 83 degrees of freedom
(depending on the model used).

\begin{table}
\caption{{{\it RXTE}} and {{\it Swift}-XRT} observations analyzed for the 2009 outburst}
\scriptsize
\begin{tabular}{lrrrl}
\hline
\hline
\RXTE & & & &\\
\hline
Start & End & Time & Observation IDs \\
(MJD) & (MJD) & (ks) &\\
55026.00  & 55038.29  & 183 & {\tt 94065-02-01-*}&\\          
& & & {\tt 94065-06-*-*}&\\
\hline
\hline
SWIFT & & & &\\
\hline
55036.71 & 55047.40  & 11 & {\tt 00030952014}-{\tt 00030952017}&\\
& & &  {\tt 00031455001}&\\
& & &  {\tt 00031456001}&\\
\hline
\hline
\end{tabular}
\label{tab:rxteobs}
\end{table}

\section{Results}

\subsection{X-ray lightcurve}
\begin{figure*}
\begin{center}
\rotatebox{0}{\includegraphics[width=18cm]{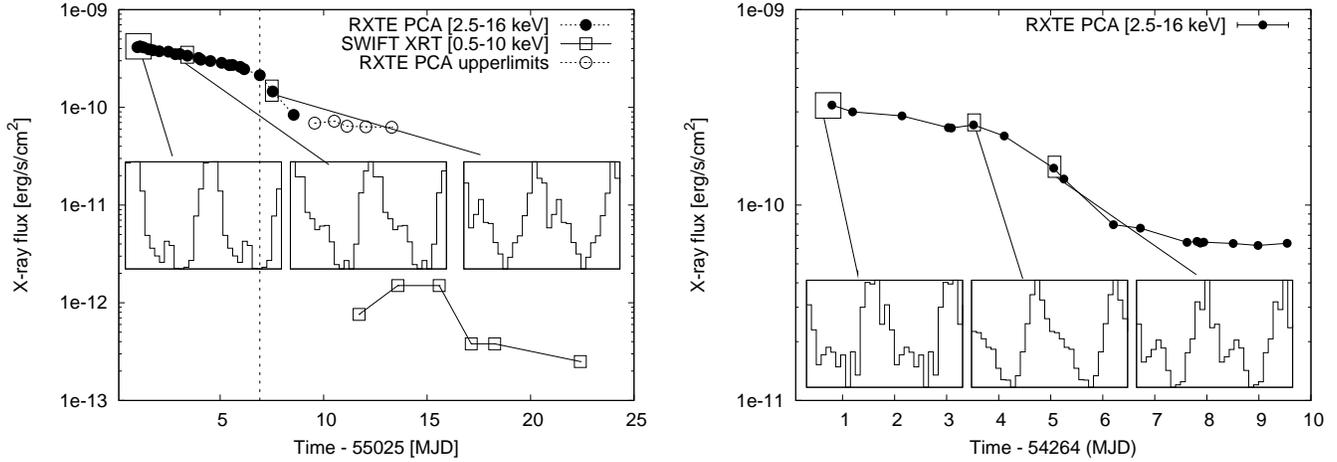}}
\end{center}
\caption{X-ray lightcurve and pulse profile evolution of 2009 (left panel) and 2007 (right panel) outbursts. 
\textbf{Left Panel:} RXTE-PCA [2.5-16 keV] and Swift XRT [0.5-10 keV] lightcurve 
of the 2009 outburst of J1756. The open circles and the open squares 
refer to 95\% confidence level upper limits. The vertical dashed line 
identifies the beginning of the fast decay. The strong contamination
of the nearby source GX 5-1 affected both the flux determinations of RXTE-PCA and Swift-XRT (see text for details). The three panels in the center show the pulse profile evolution (two cycles are plotted for clarity) obtained when folding a small chunk of lightcurve.
\textbf{Right Panel:}  RXTE-PCA [2.5-16 keV] lightcurve of the 2007 outburst, shown for comparison}
\label{lc}
\end{figure*}

We extracted the 2.5-16 keV X-ray flux from the \textit{RXTE}-PCA and
{{\it Swift}-XRT} observations. The lightcurve is shown in
Figure~\ref{lc}.  The rising portion of the lightcurve is missing
because the \textit{RXTE} monitoring started approximately five days
after the beginning of the outburst\citep{pat09}. The lightcurve
shows a slow decay for
$\sim 5$ days before entering a fast decay phase (marked with the
vertical dashed line in the figure) that brings the luminosity down to
the detection level of \textit{RXTE} on a timescale of $\sim2$
days. The slow-decay portion of the lightcurve is well fitted by an
exponential decay law with an e-folding time of $10.2\pm0.1$ days.
The fast-decay has an e-folding time of $2.7\pm 0.1$ days.  Both these
two values are longer than the timescales reported by \citet{kri07b}
for the 2007 outburst ($7.6\pm 0.4$ and $0.6\pm0.3$ days for slow and
fast decay respectively; see Figure~\ref{lc}). After only 3 days from the last
\textit{RXTE}-PCA detection, the {{\it Swift}-XRT} was unable to
detect the source.  The upper limit on the flux level at this time was
100 times smaller than that estimated from the last \textit{RXTE}-PCA
observation, and indicates that the source was probably returned
into quiescence.


\begin{figure*}
\begin{center}
\rotatebox{0}{\includegraphics[width=15.5cm]{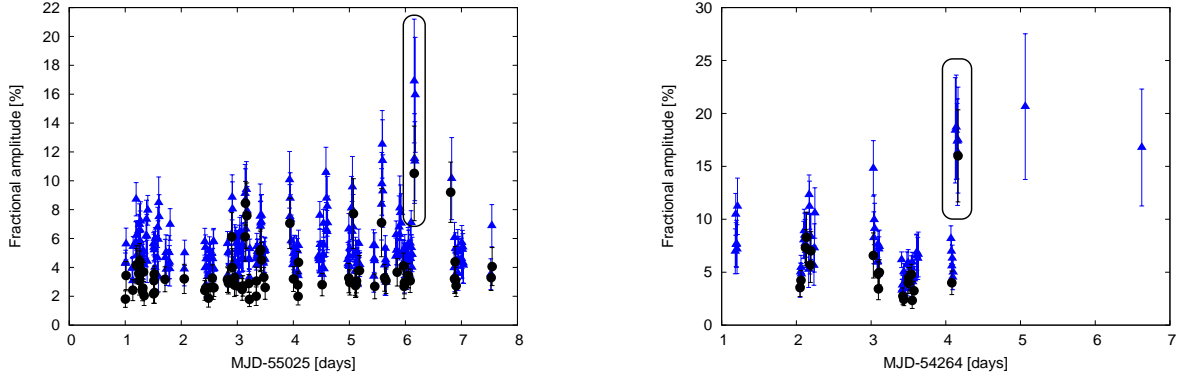}}
\end{center}
\caption{Fractional amplitude vs. time for the 2009 (left panel) and
2007 (right panel) outburst of J1756. The fundamental is indicated
with blue triangles and the first overtone with black filled
circles. The fractional amplitude varies between approximately 2 and
20\% during the two outbursts. The two circled groups of points, that
show a substantial increase in pulsed fraction, refer to the beginning
of the fast decay.}
\label{fig1}
\end{figure*}

\subsection{Color-Color Diagram}

We use the 16 s time resolution Standard 2 mode data to calculate
X-ray colours. Hard and soft colours are defined, respectively, as the
9.7-16.0 keV/6.0-9.7 keV and the 3.5-6.0 keV/2.0-3.5 keV count rate
ratios. The energy-channel conversion is done by using the
pca\_e2c\_e05v02 table provided by the RXTE team. The background was
subtracted and dead-time corrections were made. In order to correct
for the gain changes and the differences in effective area between the
PCUs themselves, we normalized our colours by the corresponding Crab
Nebula color values (see \citealt{kuu94}; \citealt{van03}) that are
closest in time but in the same RXTE gain epoch, that is, with the
same high voltage setting of the PCUs \citep{jah06}. During the 2009
observations, the source was always in the so-called island/extreme
island state, which is typical of persistent AMXPs.  The source was
also in the island/extreme island state (not shown) during the 2007
outburst, as reported by \citet{lin08}, that identified J1756 as a
typical atoll source. The colours of the source are consistent
among the two outbursts.

\subsection{Coherent timing analysis of the 2009 outburst}

 We fitted each pulse profile with a constant plus
two sinusoids: one at the nominal pulse frequency and the other at
twice the pulse frequency, representing the fundamental and the first
overtone, respectively.  We then selected only pulses with a
ratio between the amplitude and its statistical error larger than
3. Several significant first overtones were found.
We therefore analyzed separately the fundamental and the first overtone, 
as described in \citet{pat09b}.

We fitted the time of arrivals of each pulse profile with a circular
keplerian model plus a linear term representing the neutron star spin
frequency.  We used the pulsar timing program TEMPO2, which minimizes
the timing residuals between the predicted and the observed arrival
times of the pulsations \citep{hobbs}. After obtaining a converged
solution we re-iterated for a few times the whole procedure: using the
improved solution we re-folded our pulse profiles and re-fitted our
time of arrivals until convergence.  The fit is statistically
satisfactory for the fundamental (reduced $\chi^{2}=1.0$ for 251
degrees of freedom) while it is not acceptable for the first overtone
(reduced $\chi^{2}=2.8$ for 73 dof).  The reason why the first
overtone shows a bad fit relies on the presence of X-ray timing noise
in the pulse phases.  The timing noise operates on rather short
timescales, producing jumps in the timing residuals of up to 0.12
cycles on a timescale of a few hours.  This is not observed in the
fundamental, where the phase behaviour is smooth throughout the
outburst.  To partially take into account the presence of timing noise
in the first overtone, we increased the statistical errors on the
pulse phases of the first overtone by a factor 1.7. The orbital and
spin parameters of the 2009 outburst, along with upper limits on the
spin frequency derivative and the eccentricity are reported in Table
2.

We repeated the coherent timing analysis of the \textit{RXTE}-PCA
observations of the 2007 outburst of J1756 (see \citealt{kri07b} for 
a similar timing study of the 2007 outburst). Also in this outburst
several significant first overtones are detected in the pulse
profiles.  The timing solution for the 2007 outburst is reported in
Table 3.  In this case both the fundamental and the first overtone
show a low, but significant, amount of X-ray timing noise in the pulse phases. We
rescaled the pulse phase statistical errors by a factor 1.2 and 1.4
respectively, to partially take into account this effect when
estimating the timing parameter uncertainties.

The pulse fractional amplitudes, as defined in Eq.~\ref{fa}, are
reported in Figure~\ref{fig1} (note that this is the
\textit{sinusoidal fractional amplitude}, as defined in Eq.~\ref{fa},
which is a factor $\sqrt{2}$ larger than the fractional rms
amplitude).  The pulsed fraction of the fundamental fluctuates between
2 and $\sim 20\%$ during both outbursts, while the first overtone
reaches a maximum amplitude of 10-15\% and a minimum value of 2\%.
The fractional amplitudes increase substantially in correspondence
with the beginning of the \textit{fast decay} (see circled points in
Fig.~\ref{fig1} and vertical dashed line in Fig.~\ref{lc}).
Interestingly, after the beginning of the 2009 fast decay, the
fractional amplitudes return back to values comparable to the rest of
the outburst.

\begin{figure}
\begin{center}
\rotatebox{-90}{\includegraphics[width=6.5cm]{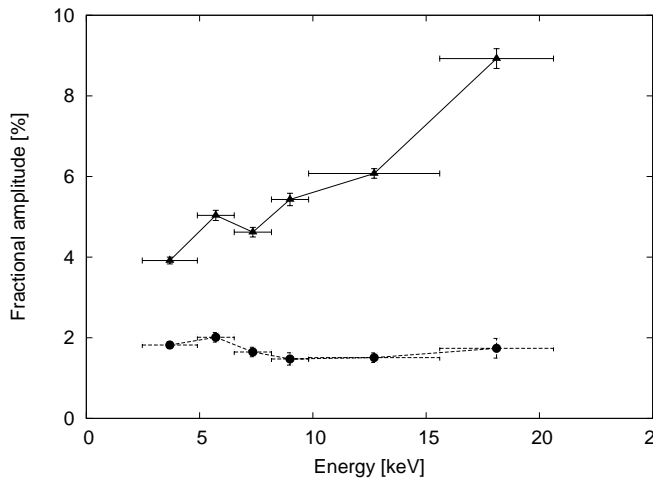}}
\end{center}
\caption{Fractional amplitude vs. energy for the 2009 outburst of
J1756. The fundamental is indicated with black triangles and the first
overtone with black filled circles. The fractional amplitude of the
fundamental increases with energy (apart from a small decrease 
around $\sim 6-7$keV).}
`\label{energy}
\end{figure}

We also studied the energy dependence of the pulsed fractions for the
2009 outburst (see Figure~\ref{energy}). The amplitude of the
fundamental increases with energy, with a slope of
$0.28\pm0.02\%\rm\,keV^{-1}$, although it is evident a slight decrease
in amplitude in the 6-7 keV energy bin. The first overtone instead
remains approximately constant with a slight decrease of amplitude
above 6 keV.


\begin{table*}
\caption{Timing solution for Swift J1756--2508 (\textbf{2009 outburst})}
\scriptsize
\begin{center}
\begin{tabular}{lllll}
\hline
\hline
Parameter & Fundamental & First overtone\\
\hline
Spin frequency $\nu$ (Hz)  & 182.06580391(2) Hz & 182.06580393(5)\\
Spin frequency derivative $\dot{\nu}$ (Hz/s) & $<3\times10^{-13}$ (95\% c.l.) & $<1.2\times10^{-12}$ (95\% c.l.)\\
Orbital period (s) & 3282.32(3) & 3282.38(6) \\
Projected semimajor axis (lt-ms) &  5.98(2) & 5.96(3)\\
Time of passage to the ascending node (MJD) & 55026.03431(3) &  55026.03425(6)\\
Eccentricity & $< 0.01$ (95\% c.l.) & $< 0.01$ (95\% c.l.) \\
$\chi^{2}/dof$ &  250.9/251  & 71.0/73$^{a}$ \\
\hline
\end{tabular}
\end{center}
$^{a}$ The value of $\chi^{2}$ refers to the fit where the pulse phase uncertainties 
have been rescaled by a factor 1.7
\label{tab:timsol09}
\end{table*}


\begin{table*}
\caption{Timing solution for Swift J1756--2508 (\textbf{2007
outburst})}

\scriptsize
\begin{center}
\begin{tabular}{lllll}
\hline
\hline
Parameter & Fundamental & First overtone\\
\hline
Spin frequency $\nu$ (Hz)  & 182.06580393(7) Hz & 182.06580378(24)\\
Spin frequency derivative $\dot{\nu}$ (Hz/s) & $<3\times10^{-12}$ (95\% c.l.) & $<1.6\times10^{-11}$ (95\% c.l.) & \\
Orbital period (s) & 3282.41(15) & 3282.27(26)\\
Projected semimajor axis (lt-ms) &  5.95(4) & 6.04(6)\\
Time of passage to the ascending node (MJD) & 54265.28087(10) & 54265.28078(14)\\
Eccentricity & $< 0.03$ (95\% c.l.) &  $< 0.03$ (95\% c.l.)\\
$\chi^{2}/dof$ &  59.94/63$^{a}$ & 13.12/14$^{b}$\\
\hline
\end{tabular}
\end{center}
$^{a}$ The value of $\chi^{2}$ refers to the fit where the pulse phase uncertainties 
have been rescaled by a factor 1.2.
$^{b}$ The value of $\chi^{2}$ refers to the fit where the pulse phase uncertainties 
have been rescaled by a factor 1.4.
\label{tab:timsol07}
\end{table*}
\subsection{Emission lines in the X-ray spectrum}

We searched for the presence of emission lines in the broad-band
spectrum of the longest RXTE observation. In this way our statistics
are the best and at the same time we are sure that the spectrum has
not changed significantly in time. 

\begin{figure}
\center
\resizebox{1\columnwidth}{!}{\rotatebox{-90}{\includegraphics{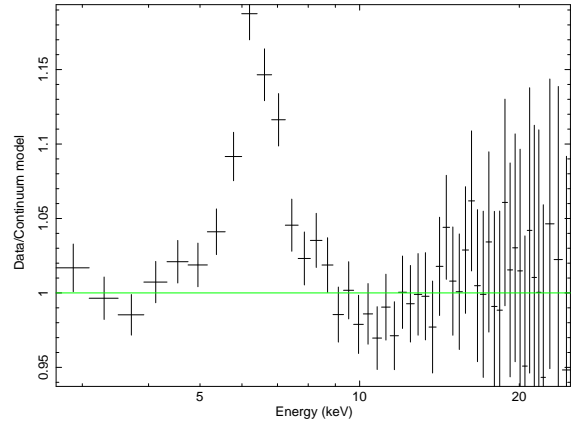}}}
\caption{Ratio of the data to the continuum model. The flux of the
  line is $(8\pm2)\times 10^{-4}$ photons cm$^{-2}$ s$^{-1}$. These dataset
  correspond to ObsID:94065-06-02-04. }
\label{fig:FE}
\end{figure}

A line is clearly detected in the 6-7 keV band, in several
occasions. For example, in Figure~\ref{fig:FE} we show the ratio of
the data to the continuum model. In this particular case, the flux of
the line is $(8\pm2)\times 10^{-4}$ photons cm$^{-2}$ s$^{-1}$.  We
also performed a similar analysis on data from the 2007 outburst; we
found several cases with significant emission lines in the same energy
range as in 2009. Although we cannot confirm the nature of the line in
the data, the detection could be suggestive of the presence of an Iron
line.  The low energy resolution of our RXTE data does not allow us to
constrain with high accuracy the energy of the line, the emissivity
index, the inner radii of the disc and/or the inclination of the
system. It is also hard to constrain the presence of a Compton
reflection bump in the X-ray spectrum because of the poor statistics
given by the very low number of counts at high energies. XMM-Newton,
Chandra or Suzaku observations are needed to verify the nature of the
line.

\section{Discussion}

\subsection{Pulse profiles}

The pulse profile shape of J1756 shows a substantial variability
during the two outbursts.  Pulse shape variability is related with the
variability of pulse phases and the variation of pulse amplitudes with
time. The pulse profiles appear to have a similar evolution in the two
outbursts, with similar profile shapes for similar outburst phases
(see Fig.~\ref{lc}). In particular, the profiles start with an almost
sinusoidal shape that evolves in time and turns into a skewed profile
during the slow decay and becomes double peaked during the fast decay
stage. Unfortunately, due to low counting statistics, it was not
possible to further follow the evolution of the pulse profiles as the
source returns to quiescence.  However, it appears intriguing the
similarity of this behaviour with that of SAX J1808.4-3658
(\citealt{har08}, \citealt{har09}), where the pulse profiles exhibit a
similar evolution of shapes between outbursts. The presence of a
stronger secondary bump toward the end of the outburst is also
interesting in light of the antipodal spot obscuration model recently
proposed by \citet{ibr09}.  The authors propose a receding inner
accretion disc as a possible cause of obscuration of the antipodal
spot, that slowly re-emerges toward the end of the outburst, when the
disc is emptying. This moves the antipodal hot spot into
the line of sight of the observer and increases the relative strength
of the first overtone over the fundamental, thus producing a stronger
secondary bump in the pulse profile.

\subsubsection{Timing noise}

The study of the 2009 pulse phases reveals a lack of timing noise in
the pulse phases of the fundamental, and a short-timescale timing
noise on the first overtone.  This appears opposite to previous
studies in which the first overtone has been considered somehow more
stable than the fundamental.

We repeated the analysis of the 2007 outburst to check the long-term 
spin and orbital evolution of J1756. 
Our orbital solution is consistent with that reported in
\citet{kri07b} (see also the erratum \citealt{kri09} for a correction
on the value of $T_{asc}$). We found also an offset of $0.32\mu$Hz from
the spin frequency reported in \citet{kri07b}. The reason of this
discrepancy is more subtle and probably has to be identified in the
difference between the timing techniques adopted in this paper and in
\citet{kri07b}.  Indeed, \citet{kri07b} did not perform an analysis of
the fundamental and first overtone separately.  As shown by
\citet{har08} and \citet{pat09b}, significant deviations from the true
timing solution arise when performing a coherent study in presence of
pulse profile variability.

The new improved set of ephemeris is not only important for the
interpretation of the X-ray data, but is crucial for understanding
the long term evolution of the neutron star spin and for any radio-search
for pulsations in this AMXP. 

\subsubsection{Pulse amplitudes}

The pulse profiles of J1756 have two remarkable features: a steep 
increase of fractional amplitudes with energy and a sudden
increase of amplitude in correspondence with the beginning 
of the fast decay. 

The energy dependence of the profiles is not new among AMXPs, as it
has been already observed and discussed in three sources: IGR
J00291+5934 \citep{fal05}, Aql X-1 \citep{cas08} and SAX J1748.9-2021
\citep{pat09c}. A possible reason for this is given by photoelectric
absorption around the rotating hot spots.  A very similar slope of the
linear relation governing the amplitude energy dependence was reported
for the intermittent source SAX J1748.9-2021 \citep{pat09c}.  However,
SAX J1748 and Aql X-1 (both intermittent pulsars) were observed to
pulsate in the soft state (\citealt{cas08}, \citealt{alt08}), while
J1756 is in the hard state (as IGR J00291), which is a more common
feature among AMXPs. This suggests that the observed increase of
pulsed amplitudes with energy is not a specific feature of
intermittent sources nor of soft or hard states.

More difficult to explain is the sudden increase of fractional
amplitude of both harmonics with the beginning of the fast decay.  
Alternate variations of pulse amplitudes after the beginning
of the fast decay were already observed in the pulsed
fractions of SAX J1808.4-3658 (see Fig. 1 in \citealt{har08}).
These increase might be connected with some fundamental 
readjustment of the accretion disc that corresponds 
to the onset of the fast decay (see \citealt{las01} for 
a review of the different outburst stages).

\subsection{Torques during outbursts}

Upper limits on any spin frequency derivative (both spin up and spin
down) can be made using the coherent timing analysis of the 2009
outburst: $|\dot{\nu}|\simless 3\times 10^{-13}\rm\,Hz\,s^{-1}$ at the
95\% confidence level.  If standard accretion torque theory is
correct, the magnetospheric radius ($r_{m}$) should decrease with the
increase of the mass accretion rate $\dot{M}$, following a power-law
$r_{m}\propto \dot{M}^{-\alpha}$ when $r_{m}<r_{co}$, with
$\alpha=2/7$ in the simplest case. The term $r_{co}$ is the the
co-rotation radius.  Assuming a source distance of 8 kpc
\citep{kri07b}, the expected mass transfer rate at the peak of the
outburst is $\dot{M}\approx 8\times 10^{-10} M_{\odot}\rm\,yr^{-1}$.
The expected spin frequency derivative is:
\begin{eqnarray}\label{torque}
\dot{\nu}&=&\frac{\dot{M}\sqrt{GMr_{m}}}{2\pi I}\simeq 1.6\times
10^{-13}\rm\,Hz\,s^{-1}\nonumber\\
&\times&\left(\frac{\dot{M}}{10^{-10}M_{\odot}\rm\,
yr^{-1}}\right)\left(\frac{\nu}{\rm\,Hz}\right)^{-1/3}\left(\frac{r_{m}}{r_{co}}\right)^{1/2}
\end{eqnarray}
 Therefore the expected maximum spin frequency derivative for J1756 
has to be of the order of $2\times 10^{-13}\rm\,Hz\,s^{-1}$, 
which is within our measured upper limits.

\subsection{Torques during quiescence}

By using the spin frequencies of 2007 and 2009 outbursts we can also
put some upper limits on the presence of a spin frequency derivative
during quiescence. The difference in frequencies between the 2007 and
2009 is 0.02$\mu$Hz, which is within the uncertainties of the
frequency determination. By propagating the uncertainties on the 2007
and 2009 spin frequencies, we can obtain an upper limit for any change
in spin frequency during quiescence: $|\Delta{\nu}|\simless 0.12\mu$Hz
(95\% confidence level) and therefore $|\dot{\nu}|\simless 2\times
10^{-15}\rm\,Hz\,s^{-1}$ (95\% c.l.).  This value is not particularly
constraining, and is a factor 4 larger than the value measured by
\citet{har08, har09} for the pulsar SAX J1808.4-3658.

By using the relativistic force-free MHD models
of pulsar magnetospheres described by \citet{s06}, we expect a magneto-dipole
torque:
\begin{equation}
N_{\rm dipole} = -\mu^2 (2\pi\nu/c)^3(1 + \sin^2\alpha)
\end{equation}
where $\mu$ is the magnetic dipole moment and $\alpha$ is
the angle between the magnetic and rotational poles.
This provides an upper limit on the magnetic dipole moment 
and therefore on the pulsar magnetic field:
\begin{eqnarray}
  \mu & < & 4.5\times10^{26}
    \left(1 + \sin^2\alpha\right)^{-1/2}
    \nonumber\\ & & \times
    \left(\frac{I}{10^{45}\ {\rm g\ cm^2}}\right)^{1/2}
    \left(\frac{\nu}{182\ {\rm Hz}}\right)^{-3/2}
    \nonumber\\ & & \times
    \left(\frac{-\dot\nu}{1.8\times10^{-15}\ {\rm Hz\ s^{-1}}}\right)^{1/2}
    \textrm{ G cm}^3\, .
  \label{eq:spindowndipole}
\end{eqnarray}
By assuming $\alpha=0$, we obtain an upper limit for the neutron star magnetic
field of $B<9\times 10^{8}$G. 
If J1756 is located close to the Galactic Center at a distance of 8 kpc, 
a lower limit for the magnetic field can be calculated by using the 
same argument as in \citet{psa99}, which assumes that the $B$ field
is strong enough to channel the accreting gas at the maximum 
mass accretion rate. By using the value $\dot{M}\approx 8\times 10^{-10}\rm\msun\,yr^{-1}$
we obtain a magnetic field range of:
\begin{equation}
0.4\times 10^{8}G\simless B\simless 9\times 10^{8}G
\end{equation}
These values are similar to those obtained for other AMXPs and 
confirm the rather narrow range of neutron star magnetic fields
in accreting systems. 

\subsection{Emission lines and X-ray spectrum}

The presence of an emission line in the 6-7 keV range suggest the
intriguing possibility of an Iron line in the broadband spectrum of
J1756. A broad relativistic Iron line has been detected only in one
AMXP until now (\citealt{pap09},\citealt{cac09}), and its detection in
J1756 can help to strongly constrain the inner accretion disc radius
and the neutron star magnetic field strength.  However, care has to be
used when claiming the detection of Iron lines in X-ray binaries,
since we do not detect a Compton reflection bump in the X-ray spectrum
and we do not have a sufficient spectral resolution to verify whether
the line is composed by a superposition of narrow lines or whether it
is a single genuinely relativistically broadened line. The reason why
we do not detect a Compton bump in our data can be due to a lack of
statistics at high energies, and a deeper observation is required to
verify the Iron line hypothesis.

\section{Conclusions}

Swift J1756.9-2508 has shown a short recurrence time of only two years
and it is the optimal candidate to study the long-term evolution
of neutron star spin and orbital parameters. 

The pulse profiles evolve in time with strong similarities between the
two outbursts and show a steep increase of pulsed fraction
with energy. With the increasing number of AMXPs exhibiting an hard
energy dependence of pulse amplitudes, it becomes more likely to 
discover new AMXPs among the
large family of the non pulsating low mass X-ray binaries. Indeed, all the
12 AMXPs discovered until now have been observed with X-ray telescopes
which are sensitive to energies below 20 keV, or that are not designed
to perform efficient coherent X-ray timing observations. With the
forthcoming X-ray telescopes like ASTROSAT, which will have a very
high effective area up to $\sim80$ keV, the detection of AMXPs with
rising pulsed fractions at hard energies could be much easier and
efficient.

The pulse phases show a low, but significant, timing noise content
and special techniques have to be applied to determine the
neutron star spin evolution. 
Thanks to the measurement of timing parameters in two outbursts, 
we can now set the first constraints on the neutron star magnetic field
($0.4\times 10^{8}G\simless B\simless 9\times 10^{8}$G) that 
makes this neutron star a typical member of the AMXP family.

\section*{Acknowledgments}
We would like to thank the \textit{RXTE} and the \textit{Swift} team 
for promptly scheduling the observations of the 2009 outburst. 
We would like to thank also Tod Strohmayer and Craig Markwardt
for useful support during the observations and Rudy Wijnands 
and Anna Watts for useful discussions.

\newcommand{\nat}{Nat}
\newcommand{\mnras}{MNRAS}
\newcommand{\aj}{AJ}
\newcommand{\pasp}{PASP}
\newcommand{\aap}{A\&A}
\newcommand{\apjl}{ApJL}
\newcommand{\apss}{ApSS}
\newcommand{\apjs}{ApJS}
\newcommand{\aaps}{AAPS}
\newcommand{\apj}{ApJ}

\end{document}